\documentclass[useAMS,usenatbib]{mn2e}
\usepackage{epsf,amsmath}
\usepackage{amssymb}
\usepackage{indentfirst}
\usepackage{array,tabularx}

\begin{document}

\title[How is Binary Radio-Pulsars with Black Holes
Population Rich?]{How is Binary Radio-Pulsars with Black Holes\\
Population Rich?}
\author[V.M. Lipunov, A.I. Bogomazov, M.K. Abubekerov]{V.M. Lipunov$^{1}$\thanks{E-mail: lipunov@xray.sai.msu.ru},  A.I.
Bogomazov$^{1}$\footnotemark[1]\thanks{E-mail:
bogomazov@xray.sai.msu.ru}, M.K.
Abubekerov$^{1}$\footnotemark[2]\thanks{E-mail: marat@sai.msu.ru}\\
$^{1}$Sternberg astronomical institute, Universitetskij prospect,
13, 119992, Moscow, Russia}

\date{Accepted by MNRAS}


\maketitle

\label{firstpage}

\begin{abstract} Using "Scenario Machine" we have carried out population
synthesis of radio pulsar with black hole binaries (BH+Psr) in
context of the most wide assumptions about star mass loss during
evolution, binary stars mass ratio distribution, kick velocity and
envelope mass lost during collapse. Our purpose is to display that
under any suppositional parameters of evolution scenario BH+Psr
population have to be abundant in Galaxy. It is shown that in the
all models including models evolved by \citet{heger2},
\citet{woosley}, \citet{heger1} expected number of the black holes
paired with radio pulsars is sufficient enough to discover such
systems within the next few years.
\end{abstract}


\section{Introduction}

\indent Binary radio pulsars with black holes (BH+Psr) revelation
would be of fundamental significance for evidence of black holes
existence and for investigation precision general relativity
effects \citep{narayan,lipunov1}. In such systems parameters of
black holes - such as mass, Kerr parameter - will be measured with
precision orders of magnitudes higher than indirect estimations of
candidates in black holes masses in X-ray/BH-candidates binaries
\citep{blandford, brumberg}. Moreover, if mutual disposition is
apt, it might be able to observe propagation of radio emission
arbitrarily near to an event horizon. First accounts of possible
number of BH+Psr binaries conducted ten years ago displayed the
systems might be observed by the modern radio-astronomical
instruments \citep{lipunov1}.

However total observed radio pulsars number have increased twice
for the last 10 years and ran up to $N_{obs}^{pul}\approx 1500$
but none of them paired with black hole
\citep{manchester0,taylor,manchester,lewandowski,psrcat}. Besides
during these 10 years our conceptions of evolution of stars which
are able to produce black holes appreciably change. In particular,
considerations in favour of greater mass loss for these stars were
obtained, detailed numerical computations considering new factors
appeared \citep{heger2,woosley,heger1}.

We have carried out population synthesis of binary stars using
"Scenario Machine". Description of it's working principles may be
found in \citep{book2}.

Since binary radio pulsars with black holes have to be generated
by massive binary stars we relied on the observable statistics of
the candidates for black holes paired with OB-stars (BH+OB).

In the paper \citep{lipunov1}, it is supposed that any black hole
paired with OB-supergiant must reveal itself as a system Cyg X-1
type. But, as it was shown by \citet{karpov}, powerful X-ray
radiation is able to originate only if accretion disk around black
hole has formed. Accordingly, in this paper, we term the Cyg X-1
type systems as the subclass of such BH+OB binaries which has
accretion disks.

Note that Cyg X-1 type systems usually are not progenitors of
BH+Psr binaries. More than 90\% of them become BH+BH or single BH
after merging of components during our calculations. The Cyg X-1
system has appreciable chance of merging in the next stage -
common envelope (CE) stage \citep{bethe}. Ordinarily, BH+Psr
binaries are originating from more wide systems which do not merge
during CE stage or come through it. At that during BH+OB stage
accretion disk does not form and observer is not able to see
bright X-ray source.

Evolutionary scenario which results in BH+Psr forming may be rough
outlined as the next. We start calculation of massive binary
system. When primary (more massive) star fills it's Roche lobe
mass transfer takes place and helium star remains instead of the
primary star. As a rule black hole forms first, it's companion -
OB-star - is sufficiently separated from BH. At that in wide
systems which can survive even after second mass transfer disk
does not form (stellar wind velocity is too high near black hole
and the moment of rotation of grasped matter is too little to form
a disk). When second component fills it's Roche lobe CE-stage
begins and only wide systems (which number is greater than number
of close binaries) produce BH+Psr (of course, anisotropic kick
plays important role in this process, it leads not only disruption
of a system, but, if kick velocity and direction are apt, it can
bound a binary during supernova explosion). In scenarios with high
mass loss rate CE stage might not to begin, because mass of
optical star is not enough greater than BH mass. Another scenario
of BH+Psr creation is possible: when neutron star forms before
black hole. At that time pulsar has long and still purely
explained history. Fortunately, part of these evolutionary tracks
is small (see below).

Note that evolutionary scenario contains quite a number of key
parameters poorly explained by theory (stellar wind magnitude,
initial mass of a star which is able to form black hole, initial
binary stars mass ratio distribution, kick velocity during
relativistic stars formation, common envelope stage efficiency,
part of mass falling into BH during collapse). Although it is
possible to reduce them significantly \citep{lipunov1,lipunov2} in
present work we suppose them as independent parameters.
Qualitative influences of these parameters on stellar evolution
scenario have been investigated more than once in previous works
\citep{jager,shore,book2,lipunov2,woosley} and it is possible to
briefly delineate them in the following way.

Stellar wind magnitude essentially influences scenario for the two
reasons. The first consists in that that spherically symmetric
wind leads increasing of component separation. The second is in
that total mass loss of a star by wind may cause change of it's
remnant type (it may produce neutron star instead black hole).

Increasing initial mass $M_{min}$ of a star which is able to form
black hole one decreases possible number of black holes due to
Salpeter power law (\ref{salpeter}).

Initial binary stars mass ratio distribution is important because
BH+Psr binaries have massive progenitors, so in case of more flat
(${\alpha_q->0}$) distribution (\ref{ratio}) probability of BH+Psr
systems birth decreases.

Anisotropic kick during compact star formation is detailed
investigated by \citet{lipunov2}. Increase of kick velocity leads
to total decay of a binary stars including relativistic companion.
It was shown that average kick velocity $<v>\approx 150-180$ km
s$^{-1}$ accords to the observable number of binary radio pulsars.
For the last few years two component kick velocity distribution
with characteristic velocities $90$ km s$^{-1}$ and $500$ km
s$^{-1}$ were suggested \citep{arzoumanian}.

Effectiveness of CE stage is described by parameter
$\alpha_{CE}=\Delta E_b/\Delta E_{orb}$, where $\Delta
E_b=E_{grav}-E_{thermal}$ is the binding energy of the ejected
envelope matter and $\Delta E_{orb}$ is the drop in systems
orbital energy during spiral-in \citep{shore}. Our population
synthesis results shows very weak dependence on CE stage
efficiency due to flat initial distribution of major semi-axis of
binaries (in logarithmic scale). If one is decreasing
$\alpha_{CE}$ close binary systems are originating from more and
more wide binaries, which number is not changing on account of
flat initial distribution \citep{shore}. We suggest
$\alpha_{CE}=0.5$ \citep{lipunov2}, that is satisfactory very
close binary systems to form even in case of a very high stellar
wind. Henceforward we do not investigate any dependencies on this
parameter.

Part of mass falling into BH during collapse is denoted as
$k_{bh}=M_{bh}/M_{preSN}$, where $M_{bh}$ is black hole mass,
$M_{preSN}$ - pre-supernova mass. It is important value because
binary can experience possible decay concerned with quantity of
mass ejected by a star during supernova explosion, i.e. the
smaller $k_{bh}$ the smaller chance a system has to survive in the
cataclysm.

We emphasize that purpose of this work is not to find optimal
model(s) of stellar evolution (another our work will be
concentrated on it), but it is to display that radio pulsar with
black hole systems number in Galaxy have to be sufficient enough
under any suppositional parameters of evolution scenarios to
observe them within the next few years. So we have obtained
multiple modelling using various appropriate parameters.

At that, as in the work \citep{lipunov1}, we were calculating
ratio of binary radio pulsars with black holes number to the total
number of pulsars (practically the last number is the number of
single pulsars). It allows to be saved from many selection effects
concerned with our lack of knowledge of average polar pattern of
the pulsars, their lifetime, magnetic field decay time,
distribution of the velocity of the pulsars and their spatial
distribution emerging at calculations of absolute number as long
as we suppose that physical parameters of radio pulsars paired
with black holes have no differences relative to average
characteristics of the pulsars of the field. This natural
suggestion is righteous because all of them have the same
progenitors - massive ($M>10M_{\odot}$) stars and, as it was shown
by our calculations, part of non-typical evolutionary tracks (when
binary radio pulsars parameters might be dissimilar to parameters
of single radio pulsars, for instance, due to effect of recycling)
in BH+Psr binaries and any other radio pulsars (binary and, of
course, single) is negligible (in most of models this quantity is
smaller than $5\%$ and never higher than $35\%$).

Since BH+Psr systems have not been observed yet we suggest that
observational limit is $\frac{BH+Psr}{Psr}\lesssim\frac{1}{1500}$,
where $1500$ is rounded number of observed radio pulsars - binary
and single \citep{psrcat}.

\section{Models description}

\indent Common parameters for all of the models are \citep{book2}:
initial mass distribution (Salpeter function) (\ref{salpeter}),
where $M_1$ - initial mass of more massive star, mass ratio
function (\ref{ratio}), where $q=M_2/M_1$ - initial component's
mass ratio and $\alpha$ takes on a values $0$ and $2$, semi-major
axis distribution (\ref{axis}), where $a$ is semi-major axis
within the limits of $10R_{\odot} < a < 10^{6}R_{\odot}$
\citep{tutukov}:

\begin{equation}
f(M)=M_1^{-2.35}, \label{salpeter}
\end{equation}

\begin{equation}
f(q)=q^{\alpha}, \label{ratio}
\end{equation}

\begin{equation}
f(a)=\frac{1}{a}, \label{axis}
\end{equation}

Kick velocity along with mass loss rate is one of the crucial and
bad fixed parameters affecting population synthesis results.
Higher kick leads to reduction of number of binary systems
containing relativistic companions \citep{lipunov2}. Although
information about kick for neutron stars may be received from the
observable binary radio pulsars statistics \citep{lyne,hansen} and
from single pulsars velocities, neutron stars kick is still under
discussion \citep{willems,murphy,podsiadlowski}.

In this work we have assumed the Maxwellian distribution of the
kick velocity $v$ during neutron star and black hole formation:

\begin{equation}
f(v)\sim \frac{v^2}{v^2_0}e^{-\frac{v^2}{v_0^2}}, \label{Maxwell}
\end{equation}
where $v_0$ is characteristic kick velocity.

Even less it is known of probable black hole kick, that is why we
vary this quantity $v_{0}^{bh}$ within the bounds of 0 and 1000
$km$ $s^{-1}$, undoubtedly exceeding observable uncertainties,
characteristic kick velocity absolute value for black holes also
depends on part of mass lost during black hole formation:

\begin{equation} v_{0}=v_{0}^{bh} \frac{M_{preSN}-M_{bh}}{M_{bh}},
\label{kicktype}
\end{equation}
where $M_{presn}$ - star mass before collapse, $M_{bh}$ - black
hole mass.

The most important characteristic - mass loss rate of optical
stars during evolution we have depicted by dint of five models -
A, B, C, Wc and Wb.

Scenario A has a weak stellar wind. Mass loss rate $\dot M$ during
main sequence (MS) stage \citep{jager} is:

\begin{equation}
\dot M\sim L/V_{\infty}, \label{mlossa1}
\end{equation}

\noindent here $L$ - star luminosity, $V_{\infty}$ - wind velocity
at infinity.

For giants we take maximum between (\ref{mlossa1}) and result
obtained by \citet{lamers}:

\begin{equation}
\dot M\sim L^{1.42}R^{0.61}/M^{0.99}, \label{mlossa2}
\end{equation}

\noindent where $R$ is stellar radius, $M$ - it's mass.

And for red supergiants we take maximum between (\ref{mlossa1})
and Reimers's formula \citep{kudritzki}:

\begin{equation}
\dot M\sim LR/M, \label{mlossa3}
\end{equation}

Mass change $\Delta M$ in wind type A during one stage is no more
than $0.1(M-M_{core})$, where $M$ is star's mass in the beginning
of a stage, $M_{core}$ - it's core mass. Mass loss during
Wolf-Rayet (WR) star stage is parametrized as $0.1\cdot M_{WR}$,
where $M_{WR}$ - maximum star mass during WR stage. We used for
calculations of stellar wind type A core masses obtained by
\citet{varshavskii, iben1, iben2}.

Scenario B uses calculations of single-star evolution by
\citet{schaller}. According to these calculations, a massive star
loses most of its mass because of the action of stellar wind, down
to $\approx 8-10M_{\odot}$ before collapse, practically
independent of it's initial mass.

In scenario C stellar evolution model is based on the results of
\citet{vanbeveren}, which reproduce most accurately the observed
galactic WR star distributions and stellar wind mass loss in
massive stars. Mass loss by a star calculations were conducted if
we used the next formula:

\begin{equation}
\Delta M=(M-M_{core}), \label{m_A}
\end{equation}

\noindent where $M_{core}$ is stellar core mass
($\ref{m_core_C}\alpha-\ref{m_core_C}\epsilon$). If maximum star's
mass (usually it is initial mass of a star, but mass transfer in
binary system is able to increase it's mass over initial value)
$M_{max}>15M_{\odot}$ mass of core in main sequence stage is
determined as ($\ref{m_core_C}\alpha$), in giant and in supergiant
stages as ($\ref{m_core_C}\beta$). In Wolf-Rayet star stage, if
$M_{WR}<2.5M_{\odot}$ and $M_{max}\le 20M_{\odot}$ it is described
as ($\ref{m_core_C}\gamma$), if $M_{WR}\ge 2.5M_{\odot}$ and
$M_{max}\le 20M_{\odot}$ as ($\ref{m_core_C}\delta$), if
$M_{max}>20M_{\odot}$ - as~($\ref{m_core_C}\epsilon$).

\begin{equation}
M_{core} = \left\{
\begin{array}{l}
1.62M_{max}^{0.83}, \qquad\qquad\qquad\qquad\qquad\;\;\,\, (\alpha) \\
10^{-3.051+4.21\lg M_{max} -0.93(\lg M_{max})^2}, \; (\beta) \\
0.83M_{WR}^{0.36}, \qquad\qquad\qquad\qquad\qquad\;\;\;\, (\gamma) \\
1.3+0.65(M_{WR}-2.4), \qquad\qquad\;\;\;\;\,\, (\delta) \\
3.03M_{max}^{0.342}, \qquad\qquad\qquad\qquad\qquad\;\;\, (\epsilon) \\
\end{array}\right.
\label{m_core_C}
\end{equation}

Scenario C has high mass loss during the WR stage, it may reach
$50\%$ of a star mass or more here. Mass loss in other stages (MS,
giant, supergiant) for stars with masses higher than $15M_{\odot}$
(for less massive stars this scenario equals to type A wind) may
reach $\approx 30\%$ of mass of a star. Total mass loss $\Delta M$
during all stages always is smaller than in scenario B and greater
than in scenario A.

W model is made in two types: with moderate (Wc) and with high
(Wb) stellar wind. Pre-supernova, helium core and compact remnant
masses according to the initial star mass to calculate parameter
$k_{bh}$ were taken from \citet{woosley}, fig. 16, \citet{heger2},
fig. 2. As it should be we calculate $k_{bh}$ as the next ratio:
$M_{bh}/M_{preSN}$, where $M_{preSN}$ is mass of pre-supernova
star which has produced black hole with mass $M_{bh}$. We made our
calculations in assumption that model Wc has type C wind, for
model Wb we used type B wind. This suggestion is quite right
because \citet{schaller} made his calculations using stellar wind
obtained by \citet{nieu} (this type of wind was used by
\citet{woosley}, \citet{heger2} as high mass loss type),
\citet{vanbeveren} - including mass loss rate by \citet{hamann}
(this type of wind was used by \citet{woosley}, \citet{heger2} as
reduced mass loss type).

Parameter $k_{bh}$ in A, B and C model is an adjusted constant
value for all supernova explosions, in Wc and Wb models it is
various quantity dependent on initial mass of a star which is able
to produce black hole.

Finally it is necessary to introduce minimal mass of black hole
$M_{min}$ (i.e. minimal pre-supernova star mass which is able to
form black hole multiplied by $k_{bh}$) which have been used for
calculations in models A, B and C. We vary this parameter within
very wide bounds: ${2.5 M_{\odot} \le M_{min} \le 10 M_{\odot}}$.

\section{Observational foundation}

\indent To estimate probable number of binary pulsars with black
holes there is a need to be guided by observable quantity of black
holes candidates in our galaxy. From this viewpoint the nearest
black hole paired with radio pulsar relation is Cyg X-1 type
system - black hole paired with blue supergiant. Namely these
binaries (BH+OB) are radio pulsars paired with black holes
progenitors. As it is well known we observe only the one such
source in the Galaxy for the time present - Cyg X-1. Although it
is not possible to speak about a statistics, we suggest following
\citet{paradijs} that this object is not a statistical ejection
and total number of such systems in Milky Way may reach a few.
Similar candidates existence in the Large Magellanic Cloud - LMC
X-1, LMC X-3 - convinces us of this.

As we have already marked Cyg X-1 is not simply binary consisting
of black hole and blue supergiant but it is a very close binary
system in which the regime of disk accretion on the black hole has
been realized. It is not by chance - very low stellar wind
velocity is necessary to form accretion disk \citep{book1}:

\begin{equation}
V<V_{cr}\approx
320(4\eta)^{\frac{1}{4}}m^{\frac{3}{8}}T_{10}^{-\frac{1}{4}}R_{8}^{-\frac{1}{8}}(1+\tan^2\beta)^{-\frac{1}{2}},
\label{condition}
\end{equation}

\noindent where $\eta$ - averaged over z-coordinate dynamic
viscous coefficient, $m=M_x/M_{\odot}$, $M_x$ - relativistic star
mass, $T_{10}=T/10$, $T$ - orbital period in days,
$R_{8}=R_{min}/10^{8}$cm., $R_{min}$ - minimal distance from the
compact object up to which free Keplerian motion is still
possible, $\beta$ - accretion axis incline angle with respect to
the radial direction. For black holes $R_{min}=3R_g$, where
$R_g=2GM_{bh}/c^2$ .

Inasmuch as stellar wind velocity rises with increasing of
distance $R$ from the normal star approximately as follow
\citep{jager}:

\begin{equation}
V = V_{\infty} (1 - R/a)^{1/2}, \label{wvel}
\end{equation}

\noindent where $a$ is characteristic radius of the star,
accretion disk does not form in the most cases. Actually,
velocities of stellar winds are not precisely measured
\citep{jager}. But (\ref{wvel}) is quite good approximation for
this work. A spherically symmetric accretion could not set a
bright source on the sky \citep{karpov}. Thus in the present
article we everywhere assume that the system Cyg X-1 type is a
very close binary including blue supergiant with mass higher than
10 masses of the Sun in which disk accretion regime has been
realized by the data (\ref{condition}).

Let us estimate (\ref{condition}) and (\ref{wvel}) for the Cyg
X-1. \citet{davis} obtained $V_{\infty}=2300\pm 400$ km s$^{-1}$
for this system. It gives lower limit for wind velocity near black
hole $V\approx 1240$ km s$^{-1}$ (\ref{wvel}). Parameters in
(\ref{condition}) and (\ref{wvel}) for Cyg X-1 has the next values
\citep{abubekerov} - $R/a\approx 0.57$, $T_{10}=0.56$, $m=10$,
$R_8=0.1$, $\tan \beta \approx 0.1$, also we suppose that
$(4\eta)^{\frac{1}{4}}\approx 1$. So, critical velocity
${V_{crit}\approx 1170}$ km s$^{-1}$. Taking into account that
observational data have great uncertainties about $V_{\infty}$ and
(\ref{condition}) has approximate character ($V<V_{cr}\approx
...$) this coincidence can be considered as quite good. Hence our
theoretical estimation (\ref{condition}) and (\ref{wvel}) is in
approximate agreement with observational data about accretion disk
forming conditions in the Cyg X-1 system.

We especially note that the Cyg X-1 is not the direct forerunner
of the binary radio pulsar with black hole, most likely this
system will merge after Roche lobe infill and common envelope
stage \citep{bethe}. However, it is clear that Cyg X-1 type
systems are very similar to BH+Psr progenitors which becomes
apparent in correlation between their abundances during population
synthesis (see below).

During population synthesis we were picking out only the pulsars
with black holes having observable orbital periods (less than 10
years) \citep{lamb, book3, johnston, psrcat}: it is may be hard to
find larger periods and the most known binary radio pulsar orbital
period is less than 4 years.

\section{Results}

\begin{figure*}
\vspace*{0pt}
\epsfbox{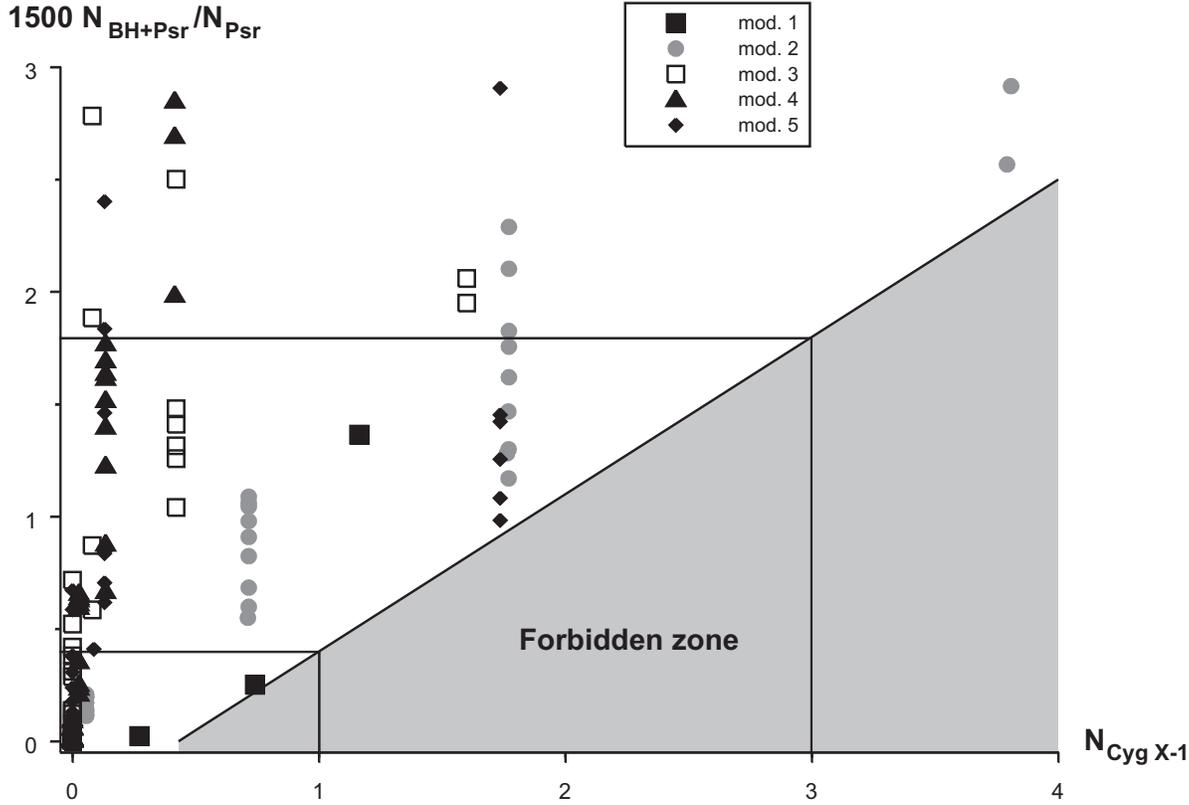}%
\caption{ Dependence between BH+Psr binaries and Cyg X-1 type
systems quantities. Black squares (mod.1) denote the model
calculated with $k_{bh}=0.45$, minimal black hole mass
$M_{min}=7M_{\odot}$, mass loss type A and uniform initial mass
ratio distribution ($\alpha_q=0$). Characteristic kick velocity
$v_0$ for neutron stars and parameter $v_0^{bh}$ for black holes
take on a values 0, 180, 360, 720 km s$^{-1}$. Number of BH+Psr
systems if $v_0=0$ and $v_0^{bh}=0$ is much more than 3, so this
point is not presented. Models in which characteristic kick
velocity for neutron stars and parameter $v_0^{bh}$ for black
holes take on a values 720, 360, 180 km s$^{-1}$ are shown
correspondingly in order of increasing of Cyg X-1 type systems
number. Grey circles (mod.2) depict models with stellar wind type
A, uniform initial mass ratio distribution ($\alpha_q=0$),
characteristic kick velocity $v_0=180$ km s$^{-1}$ for neutron
stars and parameter $v_0^{bh}=180$ km s$^{-1}$ for black holes.
During calculations quantity $k_{bh}$ has been varied within the
limits 0.1 and 1 every 0.1, black hole minimal mass takes on a
values 2.5, 3.0, 4.0, ..., 9.0, 10.0 $M_{\odot}$. Circles in the
Figure 1 evidently group in some vertical lines. Each line
corresponds to one value of $k_{bh}$, minimal black hole masses
change along lines. Cyg X-1 systems usually are products of
evolution of very massive binaries, so difference between their
number in case of $M_{min}=10M_{\odot}$ and in case of
$M_{min}=2.5M_{\odot}$ is negligible. Otherwise number of binary
radio pulsars with black holes has strong dependence on minimal
black hole mass - the more $M_{min}$ the less BH+Psr systems
number. So the bottom-up sequence order of points in lines which
depict models with various $k_{bh}$ is: $M_{min}=10, 9, ... , 4,
3, 2.5M_{\odot}$. Parameter $k_{bh}$ influences on Cyg X-1 and
BH+Psr numbers both, the more $k_{bh}$, the more number of
binaries including relativistic companion. Note that in many cases
quantities of Cyg X-1 or BH+Psr systems are higher than limits in
Figure 1. That is why in this figure presented only that models in
which $k_{bh}\le 0.6$ (for mod.2) and not all points in vertical
lines. So in mod.2 vertical lines correspond (in order of
decreasing $N_{Cyg X-1}$) to the next $k_{bh}$ values: 0.6, 0.5,
0.4, 0.3. Models with $k_{bh}=0.1,0.2$ has no Cyg X-1 type
binaries and are merged with many other models without such
systems in this figure. Open squares (mod.3) describe models with
stellar wind type C and quadratic initial mass ratio distribution
($\alpha_q=2$). Quantity $k_{bh}$, minimal black hole mass
$M_{min}$, characteristic kick velocity for neutron stars $v_0$
and parameter $v_0^{bh}$ for black holes in mod.3, mod.4 and mod.5
take on the same values as in mod.2. Also in these models as in
mod.2 points group into some vertical lines corresponding their
$k_{bh}$, minimal BH mass change along lines. Maximum $k_{bh}$ for
the presented models (mod.3) is 0.7. Black triangles (mod.4) mark
models with stellar wind type A and quadratic initial mass ratio
distribution ($\alpha_q=2$). Maximum $k_{bh}$ for the presented
models is 0.6. Black diamond formations (mod.5) designate models
with stellar wind type C and uniform initial mass ratio
distribution ($\alpha_q=0$). Maximum $k_{bh}$ for the presented
models is 0.6. "Forbidden zone" is highlighted area where there
are no any models with conceivable appropriate parameters of
stellar evolution.}
\end{figure*}

\indent The radio pulsar with black hole binaries quantity
depending on Cyg X-1 number $N_{Cyg X-1}$ numerical modelling
results are presented in the two figures. Note that we calculate
number of BH+Psr systems as the next ratio -- $1500\cdot
N_{BH+Psr}/N_{Psr}$, where 1500 is rounded number of known radio
pulsars, $N_{BH+Psr}$ is the number of BH+Psr binaries and
$N_{Psr}$ is the total number of radio pulsars appeared during
population synthesis.

There were no Cyg X-1 type systems in all of the models with type
B wind including Wb model, so all of these scenarios must be
declined right away because of we observe the object and do not
regard it as a statistical ejection.

In the Figure 1 models for the mass loss scenarios A and C at a
very wide changing of all of the unknown parameters are shown.
Since we know only one Cyg X-1 type X-ray source (namely Cyg X-1)
and purpose of this work is to display that under any
suppositional parameters BH+Psr systems have to be observable, we
have presented in Figure 1 and Figure 2 only such models in which
number of Cyg X-1 type systems is no more than 4 and number of
BH+Psr binaries is no more than 3. "Forbidden zone" in these
figures is highlighted area where there are no any models with
conceivable appropriate parameters.

Black squares (mod.1, Figure 1) denote the model calculated with
$k_{bh}=0.45$, minimal black hole mass $M_{min}=7M_{\odot}$, mass
loss type A and uniform initial mass ratio distribution
($\alpha_q=0$). Characteristic kick velocity $v_0$ for neutron
stars and parameter $v_0^{bh}$ for black holes take on a values 0,
180, 360, 720 km s$^{-1}$. We change these velocities for NS and
BH both to give a demonstration of it's influence on dependence
between BH+Psr and Cyg X-1 quantities. Number of BH+Psr systems if
$v_0=0$ and $v_0^{bh}=0$ is much more than 3, so this point is not
presented. Models in which characteristic kick velocity for
neutron stars and parameter $v_0^{bh}$ for black holes take on a
values 720, 360, 180 km s$^{-1}$ are shown in order of increasing
of Cyg X-1 type systems number (higher kick reduces of number of
binaries including relativistic companion).

Grey circles (mod.2, Figure 1) depict models with stellar wind
type A, uniform initial mass ratio distribution ($\alpha_q=0$),
characteristic kick velocity $v_0=180$ km s$^{-1}$ for neutron
stars and parameter $v_0^{bh}=180$ km s$^{-1}$ for black holes.
During calculations quantity $k_{bh}$ has been varied within the
limits 0.1 and 1 every 0.1, black hole minimal mass takes on a
values 2.5, 3.0, 4.0, ..., 9.0, 10.0 $M_{\odot}$. Circles in the
Figure 1 evidently group in some vertical lines. Each line
corresponds to one value of $k_{bh}$, minimal black hole masses
change along lines. Cyg X-1 type systems usually are products of
evolution of very massive binaries, so difference between their
number in case of $M_{min}=10M_{\odot}$ and in case of
$M_{min}=2.5M_{\odot}$ is negligible. Otherwise number of binary
radio pulsars with black holes has strong dependence on minimal
black hole mass - the more $M_{min}$ the less BH+Psr systems
number. So the bottom-up sequence order of points in lines which
depict models with various $k_{bh}$ is: $M_{min}=10, 9, ... , 4,
3, 2.5M_{\odot}$. Parameter $k_{bh}$ influences on Cyg X-1 and
BH+Psr numbers both, the more $k_{bh}$, the more number of
binaries including relativistic companion. Note that in many cases
quantities of Cyg X-1 or BH+Psr systems are higher than limits in
Figure 1. That is why in this figure presented only that models in
which $k_{bh}\le 0.6$ and not all points in vertical lines. So for
mod.2 vertical lines correspond (in order of decreasing $N_{Cyg
X-1}$) to the next $k_{bh}$ values: 0.6, 0.5, 0.4, 0.3. Models
with $k_{bh}=0.1,0.2$ has no Cyg X-1 type binaries and are merged
with many other models without such systems in the Figure 1.

Open squares (mod.3, Figure 1) describe models with stellar wind
type C and quadratic initial mass ratio distribution
($\alpha_q=2$). Quantity $k_{bh}$, minimal black hole mass
$M_{min}$, characteristic kick velocity for neutron stars $v_0$
and parameter $v_0^{bh}$ for black holes take on the same values
as in mod.2. Also as in mod.2 points group into some vertical
lines corresponding their $k_{bh}$, minimal BH mass change along
lines. Maximum $k_{bh}$ for the presented models is 0.7.

Black triangles (mod.4, Figure 1) mark models with stellar wind
type A and quadratic initial mass ratio distribution
($\alpha_q=2$). Quantity $k_{bh}$, minimal black hole mass
$M_{min}$, characteristic kick velocity for neutron stars $v_0$
and parameter $v_0^{bh}$ for black holes take on the same values
as in mod.2. Also as in mod.2 points group into some vertical
lines corresponding their $k_{bh}$, minimal BH mass change along
lines. Maximum $k_{bh}$ for the presented models is 0.6.

Black diamond formations (mod.5, Figure 1) designate models with
stellar wind type C and uniform initial mass ratio
distribution($\alpha_q=0$). Quantity $k_{bh}$, minimal black hole
mass $M_{min}$, characteristic kick velocity for neutron stars
$v_0$ and parameter $v_0^{bh}$ for black holes take on the same
values as in mod.2. Also as in mod.2 points group into some
vertical lines corresponding their $k_{bh}$, minimal BH mass
change along lines. Maximum $k_{bh}$ for the presented models is
0.6.

As one can see from the Figure 1 in the value area containing at
least one black hole candidate being in Cyg X-1 type system all of
the models are complying with the condition $1500\cdot
(N_{BH+Psr})/N_{Psr}\gtrsim 0.4$. In zone which has limits $1\le
N_{Cyg X-1}\le 3$ lower limit of number of binary radio pulsars
with black holes is between $\approx 0.4$ and $\approx 1.75$ in
all models.

\begin{figure}
\vspace{0pt}\epsfbox{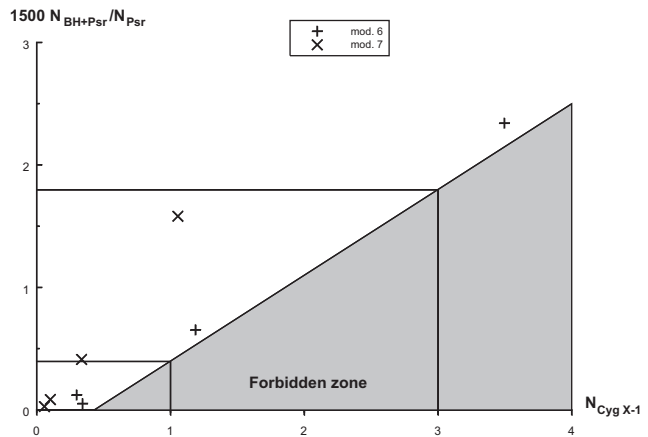}\caption{ Dependence between
BH+Psr binaries and Cyg X-1 type systems quantities (Wc model).
Upright crosses (mod.6) describe model with uniform initial mass
ratio distribution ($\alpha_q=0$). Crossbucks (mod.7) denote model
with quadratic initial mass ratio distribution ($\alpha_q=2$).
During calculations characteristic kick velocity $v_0$ for neutron
stars and parameter $v_0^{bh}$ for black holes take on a values 0,
180, 360, 720, 1000 km s$^{-1}$ in both cases. If $v_0=0$ and
$v_0^{bh}=0$ quantity of Cyg X-1 systems and BH+Psr binaries are
more than limits of the Figure 2 and appropriate point is not
presented. So upright crosses depict mod.6 and crossbucks depict
mod.7 in which characteristic kick velocity for neutron stars and
parameter $v_0^{bh}$ for black holes take on a values 180, 360,
720, 1000 km s$^{-1}$ are shown correspondingly in order of
decreasing of Cyg X-1 type systems number. "Forbidden zone" is
highlighted area where there are no any models with conceivable
appropriate parameters of stellar evolution.}
\end{figure}

In the Figure 2 two kinds of Wc models are shown (Wb model is
declined because it has no Cyg X-1 type systems).

Upright crosses (mod.6, Figure 2) describe model with uniform
initial mass ratio distribution ($\alpha_q=0$). During
calculations characteristic kick velocity $v_0$ for neutron stars
and parameter $v_0^{bh}$ for black holes take on a values 0, 180,
360, 720, 1000 km s$^{-1}$. If $v_0=0$ and $v_0^{bh}=0$ quantity
of Cyg X-1 systems and BH+Psr binaries are more than limits of the
Figure 2 and appropriate point is not presented. So upright
crosses depict mod.6 and crossbucks depict mod.7 in which
characteristic kick velocity for neutron stars and parameter
$v_0^{bh}$ for black holes take on a values 180, 360, 720, 1000 km
s$^{-1}$ are shown correspondingly in order of decreasing of Cyg
X-1 type systems number.

Crossbucks (mod.7) denote model with quadratic initial mass ratio
distribution ($\alpha_q=2$). During calculations characteristic
kick velocity $v_0$ for neutron stars and parameter $v_0^{bh}$ for
black holes take on a values 0, 180, 360, 720, 1000 km s$^{-1}$.
If $v_0=0$ and $v_0^{bh}=0$ quantity of Cyg X-1 systems and BH+Psr
binaries are more than limits of the Figure 2 and appropriate
point is not presented. So upright crosses depict mod.6 and
crossbucks depict mod.7 in which characteristic kick velocity for
neutron stars and parameter $v_0^{bh}$ for black holes take on a
values 180, 360, 720, 1000 km s$^{-1}$ are shown correspondingly
in order of decreasing of Cyg X-1 type systems number.

It is evidently from the Figure 2 that in the value area
containing at least one black hole candidate being in Cyg X-1 type
system all of the models are complying with a condition $1500\cdot
(N_{BH+Psr})/N_{Psr}\gtrsim 0.4$. In zone which has limits $1\le
N_{Cyg X-1}\le 3$ lower limit of number of binary radio pulsars
with black holes is between $\approx 0.4$ and $\approx 1.75$ in
all models.

\section{Conclusions}

\begin{figure}
\vspace{0pt}\epsfbox{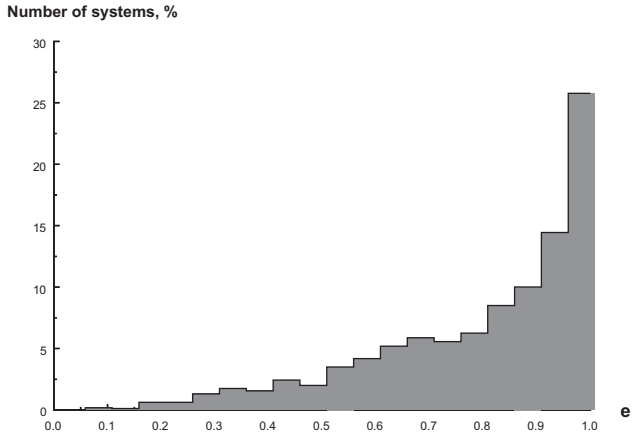}\caption{Eccentricity
distribution of BH+Psr binaries in Wc model with quadratic initial
mass ratio distribution ($\alpha_q=2$), characteristic kick
velocity for neutron stars $v_0$ and parameter $v_0^{bh}$ for
black holes take on a value 180 km s$^{-1}$.}
\end{figure}

\begin{figure}
\vspace{0pt}\epsfbox{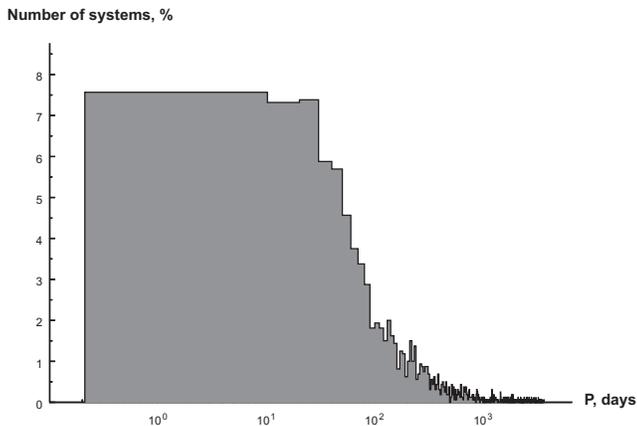}\caption{Orbital period
distribution of BH+Psr binaries in Wc model with quadratic initial
mass ratio distribution ($\alpha_q=2$), characteristic kick
velocity for neutron stars $v_0$ and parameter $v_0^{bh}$ for
black holes take on a value 180 km s$^{-1}$.}
\end{figure}

\indent \citet{lipunov1} made a conclusion that observable by
modern techniques BH+Psr binaries have to be in Galaxy. Despite
new theories concerning evolution of massive stars appearance we
confirm this conclusion about a possibility to discover binary
radio pulsars with black holes. We suggest that expected value of
pulsars paired with black holes comparative abundance may be found
within the limits $0.4\lesssim \frac{N_{BH+Psr}}{N_{Psr}} \cdot
1500 \lesssim 2$. We also confirm the conclusion made by
\citet{lipunov1} about high eccentricity of such systems (Figure
3) and about sufficient number of system close enough to observe
(Figure 4). Distribution of eccentricities shows evident peak at
$e\approx 1$ that is the consequence of mass loss and kick during
second supernova explosion \citep{kornilov}.

Our calculations are important for estimation of merging of BH+BH
and BH+NS systems. Results of \citet{panchenko} shows that merging
rate can increase by $\approx 5-7$ times with respect to previous
computations made by \citet{lipunov2}. New estimations of merging
rate of relativistic systems will be carried out in one of our
future work.

In closing we emphasize again that results of this work does not
depend on optimal evolutionary scenario parameters: radio pulsars
paired with black holes have to be in Galaxy and might be
discovered within the next few years. We accentuate that relative
number of BH+Psr systems to the total number of pulsars have been
calculated in this work and most of the pulsars do not experience
recycling effect. Radio pulsars in BH+Psr binaries originated in
our calculations have no differences with more than 90\% of radio
pulsars which have already been observed.

\section*{Acknowledgements}
\indent We thanks M.E. Prokhorov, A.G. Kuranov, S.B. Popov, I.E.
Panchenko, K.A. Postnov for their help during population synthesis
execution, S.I. Blinnikov for the important discussions and
anonymous referee for valuable notes. The work was supported by
the Russian Foundation for Basic Research (project 04-02-16 411).

\begin{table*}
 \centering
 \begin{minipage}{100mm}
  \caption{Possible evolutionary track which produces BH+Psr.}
  \begin{tabular}{@{}llrrrrc@{}}
  \hline
        &  &  &  &  &  & \\
     Stage of & Stage of& $M_1/M_{\odot}$ & $M_2/M_{\odot}$ & $a/R_{\odot}$ & Time & CE-stage\\
     primary & secondary &  &  &  &  $10^6$yr & \\
     star (1) &  star (2) &  &  &  &  & \\
 \hline
 MS & MS & 69.5 & 5.1 & 700 & 0.0 & -- \\
 SG & MS & 66.4 & 5.0 & 730 & 3.3 & -- \\
 Rov & MS & 63.5 & 5.0 & 760 & $\approx 3.6$ & + \\
 WR & Rov & 37.9 & 5.1 & 13 & $\approx 3.6$ & -- \\
 SN & & & & & & \\
 BH & Rov & 16.4 & 4.9 & 410 & 3.7 & -- \\
 BH & WR & 16.4 & 4.9 & 14 & 4.9 & -- \\
 & SN & & & & & \\
 BH & Psr & 16.4 & 1.34 & 21 & 26.3 & -- \\
\hline \label{bhpsr}
\end{tabular}
\end{minipage}
\end{table*}

\begin{table*}
 \centering
 \begin{minipage}{100mm}
  \caption{Possible evolutionary track which produces Psr+BH.}
  \begin{tabular}{@{}llrrrrc@{}}
  \hline
        &  &  &  &  &  & \\
     Stage of & Stage of& $M_1/M_{\odot}$ & $M_2/M_{\odot}$ & $a/R_{\odot}$ & Time & CE-stage\\
     primary & secondary &  &  &  &  $10^6$yr & \\
     star (1) &  star (2) &  &  &  &  & \\
 \hline
 MS & MS & 23.9 & 11.5 & 450 & 0.0 & -- \\
 SG & MS & 22.4 & 11.4 & 480 & 6.3 & -- \\
 Rov & MS & 21.0 & 11.4 & 500 & $\approx 6.9$ & -- \\
 WR & MS & 8.5 & 23.9 & 690 & $\approx 6.9$ & -- \\
 SN & & & & & & \\
 Psr & MS & 1.34 & 23.9 & 790 & 7.1 & -- \\
 Ej & SG & 1.34 & 22.3 & 830 & 13.0 & -- \\
 & SN & & & & & \\
 Psr & BH & 1.34 & 9.0 & 80 & 14.0 & -- \\
\hline \label{psrbh}
\end{tabular}
\end{minipage}
\end{table*}

\section*{Appendix: possible evolutionary tracks which produce BH+Psr binaries}

We have presented possible evolutionary track which produces
BH+Psr binary in Table \ref{bhpsr} and possible evolutionary track
which produces Psr+BH system in Table \ref{psrbh}. Note that they
do not depict all possibilities and evolution of concrete binary
strongly depends on evolutionary scenario. In both cases we used
stellar wind type A, parameter $k_{bh}=0.43$.

Marks in the tables \ref{bhpsr} and \ref{psrbh} depict the next
stages: MS - main sequence stage, SG - super-giant stage, Rov -
Roche lobe overflow stage, WR - Wolf-Rayet star stage, BH - black
hole, Psr - radio pulsar, Ej - ejecting neutron star which does
not appear itself as radio pulsar due to free-free absorbtion of
radio emission in component's stellar wind (detailed nomenclature
of neutron stars may be found in book written by \citet{book1}),
SN - supernova explosion.

First and second columns in each table contain information about
primary (more massive) and secondary companion current stage
correspondingly, third and fourth - about their masses, fifth
contains value of major semi-axis of the system, in sixth we have
presented time elapsed from the moment of binary system birth.
Seventh column answer the question - is the system in common
envelope stage or not?

Tracks which produce neutron star before black hole can lead to
so-called "recycled" radio pulsars forming. In general, we
calculate evolution of rotation of a neutron star and can include
these pulsars into consideration. Nevertheless, in this case
results of our work would be dependent on many extra assumptions:
distribution of neutron stars on their magnetic field, magnetic
field dissipation time, influence of accretion on neutron star's
magnetic field, etc. We prefer not to use any additional
hypothesis since part of such tracks is negligible (see
Introduction).

\end{document}